\documentclass[aps,pra,twocolumn,showpacs]{revtex4}
\usepackage{graphicx}
\usepackage{amssymb}
\usepackage{CJK}
\newcommand{\leftexp}[2]{{\vphantom{#2}}^{#1}{#2}}

\begin{document}
\begin{CJK*}{GBK}{song}
\title{Realizing flexible two-qubit controlled phase gate with a hybrid solid-state system}

\author{ Feng-yang Zhang\footnote{Email:zhangfy@mail.dlut.edu.cn}, Ying Shi, Chong Li,\footnote{Email: lichong@dlut.edu.cn}
and He-shan Song}
 \affiliation{School of Physics and Optoelectronic Technology,
Dalian University of Technology, Dalian $116024$ China}

\date{\today}

\begin{abstract}
We propose a theoretical scheme for realizing flexible two-qubit
controlled phase gate. A transmission line resonator is used to
induce the coupling between nitrogen-vacancy (N-V) in diamond and
superconducting qubit. The N-V center acts as control qubit and the
superconducting qubit as target qubit. Through adjusting external
flux, we obtain desired coupling between random superconducting
qubit and transmission line resonator. Moreover, our protocol might
be implemented via the current experimental technology.
\end{abstract}
  \maketitle
\newpage

Quantum computer is a magical machine, which is more efficient than
classical computers\cite{a,b,c} in solving hard computational
problems . Therefore, how to successfully construct quantum computer
is still an open issue. The core element of quantum computer is
quantum logic gate. It is well known that a quantum-computing
network can be decomposed into one qubit rotations gate and
two-qubit gates\cite{d}. So far, many theoretical and experimental
proposals demonstration two-qubit gates in different physical
system, for example, two-qubit controlled phase gates have been
experimental realized in cavity QED\cite{e}, quantum dot\cite{f},
NMR\cite{g}, ion traps system\cite{h}, respectively. Quantum
information processing (QIP) need repeated implement quantum gate
operations. However, the decoherence of system is the main stumbling
block for QIP. Seeking the nice candidate for quantum information
carrier is still a popular study field.

Due to the sufficient long electronic spin lifetime and the
possibility of coherent manipulation at room temperature\cite{i1},
the N-V center in diamond\cite{i} consisting of a substitutional
nitrogen atom and an adjacent vacancy,  provides an arena to study
various macroscopic quantum phenomena and acts as a perfect
candidate toward QIP\cite{j}. The two-qubit conditional quantum gate
has been realized in experiment\cite{n}, then the theoretical scheme
of achieving multiqubit conditional phase gate was proposed in
Ref.\cite{o}. Also, the coherent coupling between superconducting
qubit and N-V center in diamond has already been achieved\cite{p,q}.

In this paper, we propose an alternative scheme to realize flexible
two-qubit controlled phase gate by a transmission line resonator
which induces the interaction between N-V center in diamond and
random superconducting qubit. The meaning of flexible two-qubit
controlled phase gate is a controlling qubit to control random
desired target qubit. Because of the long coherence time of N-V
center, it can act as controlling qubit (first qubit), and the
superconducting qubit acts as target qubit (second qubit). All the
coupling (decoupling) between target qubits and the transmission
line resonator can be controlled by adjusting its external flux. So
the expectative two-qubit phase gate can be constructed. As shown
below, this proposal has the following advantages: (i) the coupling
technologies between the transmission line resonator and the
superconducting qubit\cite{q1}, the N-V center\cite{ad} are rather
mature in current experiments; (ii) compared with the conventional
scheme\cite{q2}, there is no need to adjust the qubit's frequency
and the transmission line's frequency throughout the entire
operation; (iii) the control qubit has relative long decoherence
time and the target qubit has flexible controllability.

\begin{figure}
  % Requires \usepackage{graphicx}
  \includegraphics[scale=0.5]{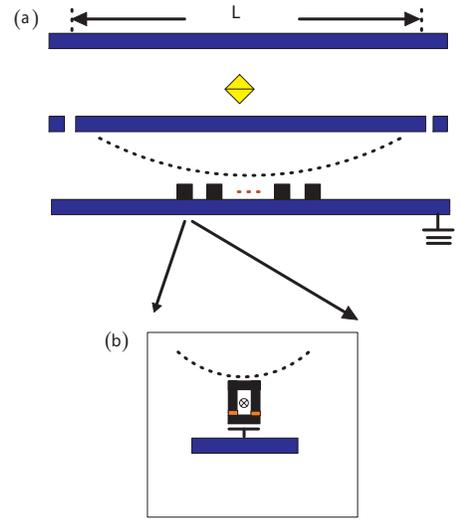}\\
  \caption{(Color online) (a) Schematic diagram of the random superconducting qubit (black frame) coupled to N-V center (yellow lozenge)
  mediated by the transmission line resonator. (b) Detailed schematic of a
charge qubit placed in antinode of the high-quality single-mode
quantum transmission line resonator.}
\end{figure}

We consider the coupling of N-V center in diamond and random
superconducting qubit induced by transmission line resonator (see
the Fig.1).

The typical superconducting qubits include charge qubit, flux qubit,
and phase qubit. Any two-level system of these qubits are commonly
described by a general Hamiltonian with Pauli matrix $\sigma_{x}$
and $\sigma_{z}$\cite{r}. In this paper, we choose charge qubit as
an example, the Hamiltonian describes the qubit is\cite{r,s}
\begin{eqnarray}
H_{q}=-4E_{c}\left(\frac{1}{2}-n_{g}\right)\sigma_{z}-\frac{1}{2}E_{J}(\Phi)\sigma_{x},
\end{eqnarray}
where the Coulomb energy $E_{c}=e^{2}/(C_{g}+2C_{J})$, the
dimensionless gate charge $n_{g}=C_{g}V_{g}/2e$,
$E_{J}(\Phi)=E_{J0}\cos(\pi \Phi/\Phi_{0})$. Here, $C_{g}$ expresses
the gate capacitance, $V_{g}$ is the gate voltage, $\Phi$ is the
external flux through the superconducting loop, the flux quantum
$\Phi_{0}=h/2e$, and $E_{J0}$ is Josephson coupling energy. If the
qubit is assumed to be biased at the degeneracy point $n_{g}=1/2$,
the Hamiltonian reduces to
$H_{q}=-\frac{1}{2}E_{J}(\Phi)\sigma_{x}$. Also, we get the
evolution operator
\begin{equation}
U_{1}(\tau)=\exp[i\zeta\sigma_{x}\tau/2],
\end{equation}
where $\zeta=E_{J}(\Phi)$ and we take $\hbar=1$ (in entire paper).

The N-V center contains a vacancy that results in broken bonds in
the system. We utilize $C_{3v}$ group theory to analyse the level
structure of N-V center in diamond\cite{i,u}. The level structure
(see Fig.2 (a)) includes the spin triplet ground state
$\leftexp{3}{A}$, excited state $\leftexp{3}{E}$, and the spin
singlet metastable $\leftexp{1}{A}$. Also, the ground states have a
zero-field splitting $D_{gs}=2.87$ GHz between degenerated sublevels
$m_{s}=\pm1$ and $m_{s}=0$\cite{v}. We can remove the degeneracy of
spin sublevels $m_{s}=\pm1$ by external magnetic field $\textbf{B}$
(see the Fig.2 (b)). The two sublevels $m_{s}=0$ and $m_{s}=-1$ can
be coupled by a microwave field, so we can encode the qubits. Under
a frame rotating with microwave field frequency, the Hamiltonian of
the N-V center could be written as\cite{q,x}
\begin{equation}
H_{N-V}=(\omega_{0}-\omega_{r})|-1\rangle\langle-1|+\frac{\Omega}{2}(|0\rangle\langle-1|+|-1\rangle\langle0|),
\end{equation}
where $\omega_{0}=D_{gs}+\gamma|\textbf{B}|$ expresses the energy
gap of the sublevels, $\gamma$ is the gyromagnetic ration of an
electron. $\omega_{r}$ and $\Omega$ are the frequency and the Rabi
frequency of the microwave pulse with the N-V center, respectively.
If the microwave field resonates to the N-V center, i.e.
$\omega_{0}=\omega_{r}$, the Hamiltonian (3) reduces to
\begin{equation}
H_{N-V}=\frac{\Omega}{2}S_{x},
\end{equation}
where $S_{x}=|0\rangle\langle-1|+|-1\rangle\langle0|$. Also, we
obtain the evolution operator
\begin{equation}
U_{2}(\tau)=\exp[i\xi S_{x}\tau/2],
\end{equation}
where $\xi=-\Omega$.
\begin{figure}
  % Requires \usepackage{graphicx}
  \includegraphics[scale=0.6]{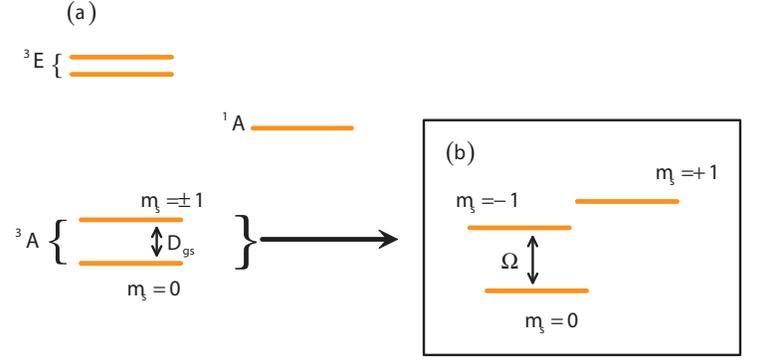}\\
  \caption{(Color online) (a) Energy level scheme of the N-V center in diamond. It includes the excited state
  $^{3}E$,  metastable $^{1}A$, and ground state $^{3}A$. $D_{gs}$ is the zero-field splitting between
the ground state sublevels $m_{s}=0$ and $m_{s}=\pm1$.
 (b) The energy-level diagram for
the ground triplet state of the electron spin within a N-V center.
We encode qubits in the subspace, the down state $m_{s}=0$ and the
up state $m_{s}=-1$, which were driven by a microwave field
$\Omega$.}
\end{figure}

In this section, we discuss how to realize flexible two-qubit phase
gate with our scheme. The total Hamiltonian of the hybrid system is
\begin{eqnarray}
H&=&\omega
a^{\dag}a-\frac{\zeta}{2}\sigma_{x}-\frac{\xi}{2}S_{x}+g(a+a^{\dag})\sigma_{x}\nonumber\\&&+G(a+a^{\dag})(S_{+}e^{i\omega_{r}t}+S_{-}e^{-i\omega_{r}t}),
\end{eqnarray}
where the first term describes the quantized transmission line
resonator with the frequency $\omega=\pi/\sqrt{LC}$, and $L (C)$ is
the total self-inductance (capacitance). $a^{\dag}$ and $a$ are the
creation and annihilation operators of the transmission line
resonator, respectively. The last two terms describe the
interactions, $g$ depicts the effective coupling strength between
the transmission line resonator and superconducting qubit, $G$
denotes the effective coupling strength between the transmission
line resonator and N-V center.

In the interaction picture, under the rotating wave approximation,
we obtain the Hamiltonian
\begin{equation}
H_{I}=g(a^{\dag}e^{i\omega t}+ae^{-i\omega
t})\sigma_{x}+G(a^{\dag}S_{-}e^{i\Delta t}+aS_{+}e^{-i\Delta t}),
\end{equation}
where $\Delta=\omega-\omega_{r}$. The external driving of the
transmission line resonator can be modeled by
\begin{equation}
H_{d}=\varepsilon(t)(a^{\dag}e^{-i\omega_{d}t}+ae^{i\omega_{d}t}),
\end{equation}
where $\varepsilon(t)$ expresses the drive amplitude and
$\omega_{d}$ is the frequency of the external drive. Under the
strong coupling limit (the coupling strength $G$ is much lager than
the resonator linewidth $\kappa$), Eq.(8) can be approximately
described by $H'_{d}=\Omega'S_{x}$, where
$\Omega'=G\varepsilon(t)/\Delta$. Thus, the whole Hamiltonian is
\cite{x1}
\begin{eqnarray}
H_{T}&=&g(a^{\dag}e^{i\omega t}+ae^{-i\omega
t})\sigma_{x}\nonumber\\&&+G(a^{\dag}S_{-}e^{i\Delta
t}+aS_{+}e^{-i\Delta t})+\Omega'S_{x}.
\end{eqnarray}
We introduce the new basis
$|\pm\rangle=(|-1\rangle\pm|0\rangle)/\sqrt{2}$, which are the
eigenstates of operator $S_{x}$ with eigenvalues $\pm1$. Under the
frame rotating transformation of the unitary operator
$U=\exp(i\Omega'S_{x})$, in the strong driving regime
$\Omega'\gg\{G,\Delta\}$, we can eliminate the fast-oscillating
terms of the Eq.(9) and obtain
\begin{eqnarray}
H_{eff}=g(a^{\dag}e^{i\omega t}+ae^{-i\omega
t})\sigma_{x}+\frac{G}{2}(a^{\dag}e^{i\Delta t}+ae^{-i\Delta
t})S_{x}.
\end{eqnarray}

The $\{\sigma_{x}S_{x}, a\sigma_{x}, a^{\dag}\sigma_{x}, aS_{x},
a^{\dag}S_{x}, I\}$ form a closed Lie algebra. Utilizing the
Wei-Norman algebraic method\cite{y}, we obtain the evolution
operator which can be written in a factorized way as
\begin{eqnarray}
U_{I}(t)&=&e^{-iA(t)\sigma_{x}S_{x}}e^{-iB(t)
a\sigma_{x}}e^{-iB^{*}(t)a^{\dag}\sigma_{x}}\nonumber\\
&&\times e^{-iC(t)aS_{x}}e^{-iC^{*}(t)a^{\dag}S_{x}}e^{-iD(t)},
\end{eqnarray}
where the time-dependent parameters
\begin{eqnarray}
A(t)=\frac{gG}{\omega}\left[\cos\Delta
t-\cos(\omega-\Delta)t\right], \nonumber\
\end{eqnarray}
\begin{eqnarray}
B(t)=\frac{ig}{\omega}(e^{-i\omega t}-1), \nonumber\
\end{eqnarray}
\begin{eqnarray}
C(t)=\frac{iG}{2\Delta}(e^{-i\Delta t}-1), \nonumber\
\end{eqnarray}
\begin{eqnarray}
D(t)=\frac{g^2}{\omega}\left[\frac{1}{i\omega}(e^{i\omega
t}-1)-t\right]+\frac{G^2}{4\Delta}\left[\frac{1}{i\Delta}(e^{i\Delta
t}-1)-t\right].
\end{eqnarray}
If the evolution time $t$ satisfies $t=2\pi n/\omega=2\pi p/\Delta$
($n,p$ are integers), the parameters $B(t)=C(t)=0$, i.e. the
interaction between transmission line resonator and N-V center, and
interaction between transmission line resonator and superconducting
qubit can be eliminated. Therefore, up to a phase factor, the
evolution operator Eq.(8) reduces as
\begin{equation}
U_{3}(n)=\exp[-iA\sigma_{x}S_{x}].
\end{equation}

After aforementioned three-step operation (2), (5) and (10), the
joint evolution operator of the system is
\begin{eqnarray}
U&=&U_{1}U_{2}U_{3} \nonumber\\
&=&\exp[i\zeta\sigma_{x}\tau/2]\times\exp[i\xi
S_{x}\tau/2]\times\exp[-iA\sigma_{x}S_{x}].
\end{eqnarray}
Under the condition $\zeta\tau/2=\xi\tau/2=A=\eta$, Eq.(11) can be
written as
\begin{equation}
U=\exp[i\eta(\sigma_{x}+S_{x}-\sigma_{x}S_{x})].
\end{equation}

In order to realize flexible two-qubit phase gate, we define the
basis of qubit $\{|gg_{j}\rangle, |ge_{j}\rangle, |eg_{j}\rangle,
|ee_{j}\rangle\}$. Here $|g\rangle=(|-1\rangle-|0\rangle)/\sqrt{2},
|e\rangle=(|-1\rangle+|0\rangle)/\sqrt{2}$,
$|g\rangle_{j}=(|\uparrow\rangle_{j}-|\downarrow\rangle_{j})/\sqrt{2}$,
$|e\rangle_{j}=(|\uparrow\rangle_{j}+|\downarrow\rangle_{j})/\sqrt{2}$,
and $|\uparrow\rangle_{j}$ $(|\downarrow\rangle_{j})$ is the
eigenstate of the operator $\sigma^{j}_{z}$ for discretionary the
superconducting qubit. If we omit the phase factor $e^{i\eta}$, and
choose $\eta=\pi/8+m\pi/2$ (for integer $m$), two-qubit controlled
phase gate can be constructed\cite{q2}
\begin{equation}
U=\left(
    \begin{array}{cccc}
      1 & 0 & 0 & 0 \\
      0 & 1 & 0 & 0 \\
      0 & 0 & 1 & 0 \\
      0 & 0 & 0 & -1 \\
    \end{array}
  \right).
\end{equation}
By adjusting the external flux $\Phi$ threading  the superconducting
qubit loop, those external flux $\Phi$ which pass through the idle
superconducting qubits are set to be $n\Phi_{0}/2$. Therefore, the
desired coupling between superconducting qubit and transmission line
resonator is achieved and the flexible two-qubit controlled phase
gate is obtained as well.

We briefly address the experimental feasibility of the proposed
scheme with the parameters already available in current experimental
setups. In 2004, Blais \emph{et al.} proposed the coupling between
superconducting qubit and transmission line\cite{z}. The size of the
superconducting qubit $1\times1\mu$m$^{2}$ has been
reported\cite{aa}. Therefore, multiple qubits can be trapped in the
space between the center conductor and the ground planes of
transmission line resonator. Later on, the transmon was proposed in
Ref.\cite{ab}, and the relaxation and dephasing time $T_{1}=1.5\mu$s
and $T_{2}=2.05\mu$s, respectively, were reported\cite{ac}.

One the other hand, in current experiments, the coupling of
transmission line resonator to an ensemble of N-V center has been
achieved\cite{ad}. Also, the dephasing and relaxation time of N-V
center are reported to be $T'_{2}=350\mu$s\cite{ae} and $T_{1}$ from
$6$ms at the room temperature up to  $28$s at the low
temperature\cite{af}, respectively. For the isotopically pure
diamond sample, the dephasing time $T'_{2}=2$ms\cite{ag}.

We choose the following parameters, the effective Josephson energy
of superconducting qubit $E_{J}=2.2\times2\pi$GHz, the frequency of
transmission line resonator $\omega=1$GHz, the coupling strength
$g=19.71$MHz\cite{ai}, the coupling constant of N-V center and
transmission line $G=11\times2\pi$MHz. Also, we take the integer
$n=1$, $m=0$ and Rabi frequency $\Omega=20\times2\pi$GHz. So,
generating two-qubit phase gate's time $t\sim6$ns is much shorter
than the decoherence time of system. The preceding rough estimates
indicate that multiple operations can be performed before
decoherence happens.

In summary, we have proposed a scheme to achieve flexible two-qubit
controlled phase gate by using solid hybrid system. Due to the N-V
center has long decoherence time at the room temperature
(especially, at the low temperature, the coherence time is much
longer), we utilize it as control qubit. The superconducting qubit
has the characteristic advantage, which owns flexible
controllability via adjusting the external parameters, therefore, it
is used as target qubit. Hence, we predict that this scheme might be
realized with current technology.

One of the authors (Feng-yang Zhang) thank Dr. Pei Pei and Jia-sen
Jin for helpful discussions. This work is supported by the
Fundamental Research Funds for the Central Universities
No.DUT10LK10, the National Science Foundation of China under Grants
No. 60703100, No. 11175033, and No. 10875020.

\end{CJK*}
\end{document}